\documentclass[12pt]{article}
\usepackage{a4,amsthm,amsfonts,latexsym
}
\swapnumbers 
\theoremstyle{plain}
\newtheorem{thm}{THEOREM}[section]

\newtheorem{lm}[thm]{LEMMA}

\newtheorem{proposition}[thm]{PROPOSITION}
\newcommand{\R}{\mathord{\mathbb R}}
\newcommand{\C}{\mathord{\mathbb C}}
\newcommand{\xoperp}{\underline{x}^\perp}
\newcommand{\yoperp}{\underline{y}^\perp}
\def\mfr#1/#2{\hbox{${{#1} \over {#2}}$}}
\def\const.{{\rm const.}}
\newcommand{\beq}{\begin{equation}}
\newcommand{\eeq}{\end{equation}}
\def\beqa{\begin{eqnarray}}
\def\eeqa{\end{eqnarray}}
\newcommand{\bsigma}{\mathord{\hbox{\boldmath $\sigma$}}}
\newcommand{\text}{\hbox}

\def\mfr#1/#2{\hbox{${{#1} \over {#2}}$}}

\begin{document}
\markboth{\scriptsize{BSY 19/11/99}}{\scriptsize{BSY 19/11/99}}
\title{\bf{Atoms in strong magnetic fields:\\The high field limit \\at 
fixed nuclear charge}}
\author{\vspace{5pt} Bernhard Baumgartner$^{1}$, Jan Philip Solovej$^{2}$, 
and Jakob 
Yngvason$^{1}$\\ 
\vspace{-4pt}\small{$1.$ Institut f\"ur Theoretische Physik, Universit\"at 
Wien}\\
\small{Boltzmanngasse 5, A-1090 Vienna, Austria}\\
\vspace{-4pt}\small{$2.$ Department of Mathematics, University of 
Copenhagen}\\
\small{Universitetsparken 5, DK-2100 Copenhagen \O, Denmark}}
\date{\small\today}
\maketitle

\begin{abstract}
Let $E(B,Z,N)$ denote the ground state energy of an atom with $N$ 
electrons and nuclear charge $Z$ in a homogeneous magnetic field $B$.  
We study the asymptotics of $E(B,Z,N)$ as $B\to \infty$ with $N$ and 
$Z$ fixed but arbitrary.  It is shown that the leading term has the 
form $(\ln B)^2 e(Z,N)$, where $e(Z,N)$ is the ground state energy of 
a system of $N$ {\em bosons} with delta interactions in {\em one} 
dimension.  This extends and refines previously known results for 
$N=1$ on the one hand, and $N,Z\to\infty$ with $B/Z^3\to\infty$ on the 
other hand.

\end{abstract}

\section{Introduction}

The effects of extremely strong magnetic fields (order of $10^9$ Gauss and 
higher) on atoms and molecules are of considerable astrophysical as well as 
mathematical interest and are far from being completely understood in spite of 
many theoretical studies since the early seventies. We refer to  \cite{LSY94} 
and 
\cite{RWHG} for a general discussion of this subject and extensive lists of 
references. 
An atom (ion) with $N$ electrons and nuclear 
charge $Z$ in a homogeneous magnetic field ${\bf B}=(0,0,B)$ is (in 
appropriate 
units) usually 
modeled by the nonrelativistic many-body Hamiltonian 
\begin{equation}H_{B,Z,N}=\sum_{i=1}^N \left(H^{(i)}_{\bf A}-{Z\over \vert 
x_{i}\vert}\right)+\sum_{i<j}^N {1\over \vert x_{i}- 
x_{j}\vert}.\end{equation}
Here $x_i\in\R^3$ are the postitions of the electrons, $i=1,\dots, N$,  
${\bf A}(x)=\mfr{1}/{2}{\bf B}\times x$ is the vector potential, and 
\beq H_{\bf A}=[({\rm i}\nabla+{\bf A}(x))\cdot \bsigma]^2\label{ha}\eeq
with $\bsigma$ the vector of Pauli spin matrices. The Hamiltonian 
$H_{B,Z,N}$ operates on the Hilbert space 
${\cal H}_{N}=\bigwedge^{N}L^2(\R^3;{\C}^{2})$ appropriate for 
Fermions of spin 1/2. In this paper we are concerned with the ground state 
energy
\beqa E(B,Z,N)&=&\inf\, {\rm spec}\, H_{B,Z,N}\nonumber \\
&=&\inf \{\langle\Psi,H_{B,Z,N}\Psi\rangle\,:\, \Psi\in 
C^{\infty}_{0}\left(\R^{3N};\C^{2^N}\right)\cap{\cal H}_N, 
\Vert\Psi\Vert_{2}=1\}
,\label{Ebzn}
\eeqa
more precisely the $B\to \infty$ asymptotics of this quantity. Such an 
asymptotic study is relevant at the field strengths prevailing on white 
dwarfs and neutron stars.

Previous investigations of the asymptotics of $E(B,Z,N)$ have 
either dealt with the case $N=1$, i.e., hydrogen-like atoms 
\cite{AHS}, or the case when $Z$ and $N$ tend to $\infty$ together with $B$
\cite{LSY94}, \cite{LSY94b}, \cite{I1}. The most complete 
rigorous treatment of the $N=1$ case so far is \cite{AHS} where the 
following $B\to\infty$ asymptotics was derived:
\begin{eqnarray} 
E(B,Z,1)/Z^2=&-&\mfr1/4[\ln(B/2)]^2+[\ln(B/2)\ln\ln(B/2)]\nonumber\\ 
&-&[(C+\ln2)\ln(B/2)]
-[\ln\ln(B/2)]^2 \nonumber\\&+&2(C-1+\ln 
2)\ln\ln(B/2)+O(1),\label{hyd}\end{eqnarray}
with a constant $C$ (Euler's constant/2). The basic results on the 
$N,Z\to\infty$ case were 
obtained in  \cite{LSY94} and \cite{LSY94b}. In particular,  
in \cite{LSY94} it was shown that if 
$N,Z\to\infty$ with $\lambda=N/Z$ fixed, and $B/Z^3\to\infty$, then
\beq E(B,Z,N)/(Z^3[\ln(B/Z^3)]^2)\to \left\{ \begin{array}{l@{\quad}l}
	 -\mfr1/4\lambda+\mfr1/8\lambda^2
-\mfr1/{48}\lambda^3 & {\rm if}\quad \lambda< 2\\ \\-\mfr1/6 & {\rm if}\quad 
\lambda\geq 2.
	\end{array}
	\right.\label{hs}\eeq
The 
fact that the right side of (\ref{hs}) decreases with increasing $N/Z$ as 
long as $N/Z<2$ shows that in the limit $Z\to\infty$, $B/Z^3\to\infty$ an 
atom can bind at least $2Z$ electrons. In 
\cite{I1} some higher order corrections to 
the leading asymptotics for the energy are discussed.

The main result of the present paper is a derivation of the leading 
term in the $B\to\infty$ asymptotics of $E(B,Z,N)$ where $Z$ and $N$ 
are {\it fixed}, but arbitrary. The precise 
statement is as follows:
\begin{thm}[High field limit of the energy]\label{1} For each fixed $Z$ and 
$N$
\beq \lim_{B\to \infty}{E(B,Z,N)\over (\ln B)^2}=e(Z,N)\label{limit}\eeq
where $e(Z,N)$ is the ground state energy of the 
Hamiltonian
\beq h_{Z,N}=\sum_{i=1}^N 
\left(-\partial^2/\partial z_{i}^2-Z\delta(z_{i})\right)+\sum_{i<j}^N
\delta (z_{i}-z_{j})\label{dh}\eeq 
of $N$ {\rm bosons} with $\delta$-interaction in one dimension, 
defined in the sense of quadratic forms 
as
\beq e(Z,N)=\inf \{\langle\Psi,h_{Z,N}\Psi\rangle\,:\, \Psi\in 
C^{\infty}_{0}(\R^{N}), \Vert\Psi\Vert_{2}=1\}.\label{ezn}\eeq
\end{thm}
It is trivial to compute $e(Z,1)=-Z^2/4$. Thus (\ref{limit}) 
generalizes the first term in the expansion (\ref{hyd}) to the case 
$N>1$. The relevance of the $\delta$-function model for the ground 
state of hydrogen in strong magnetic fields was noted already in \cite{Sp76}.

We also verify that  the mean field limit of 
$e(Z,N)$ agrees with (\ref{hs}):
\begin{thm}[Mean field limit]\label{2}If $Z,N\to\infty$ with $\lambda=N/Z$ 
fixed, then
	\beq
	e(Z,N)/Z^3\to \left\{ \begin{array}{l@{\quad}l}
	 -\mfr1/4\lambda+\mfr1/8\lambda^2
-\mfr1/{48}\lambda^3 & {\rm if}\quad \lambda< 2\\ \\-\mfr1/6 & {\rm if}\quad 
\lambda\geq 2
	\end{array}
	\right.
	\eeq
	\end{thm}

Taken together, Theorems \ref{1} and \ref{2} lead to the same high $B$, high 
$Z$ limit as Theorem 1.4
in 
\cite{LSY94}, where $Z\to\infty$ and $B/Z^3\to\infty$ simultaneously
(the \lq\lq hyper-strong\rq\rq\ limit.) 

We now describe briefly the strategy for the proof of these results and 
introduce some notation that will be used throughout. The first step in the 
proof of Theorem \ref{1} is a reduction to
the subspace ${\cal H}_{N}^0\subset
{\cal H}_{N}$ generated by wave functions in the 
lowest Landau band. Let $\Pi^0_N$ denote the projector on ${\cal H}_{N}^0$. 
(Its integral kernel is given by Eqs. (\ref{intkern})--(\ref{intkern2}) in 
Section 5. Note that $\Pi^0_N$ and ${\cal H}_{N}^0$ depend on $B$.)
Let $E^{\rm conf}(B,Z,N)$ denote the ground state energy  of 
$\Pi^0_N H_{B,Z,N}\Pi^0_N$.
It is clear that
\beq E(B,Z,N)\leq E^{\rm conf}(B,Z,N),\eeq
and by Theorem 1.2 in \cite{LSY94}, 
\beq E^{\rm conf}(B,Z,N)\leq E(B,Z,N) (1-\delta(B,Z,N))\label{conf}\eeq
where $\delta(B,Z,N)\to 0$ for $B\to\infty$ with $Z,N$ fixed. Hence it 
suffices to prove (\ref{limit}) with $E(B,Z,N)$ replaced by $E^{\rm 
conf}(B,Z,N)$.
We note in passing that (\ref{conf}) also holds for bosons. In fact, it will 
become evident in the sequel that Theorem \ref{1} is independent of the 
statistics of the particles.

To study $E^{\rm conf}(B,Z,N)$ the next step is to introduce
a Hamiltonian for the motion parallel 
to the magnetic field with the coordinates perpendicular to the 
magnetic field 
as parameters.
We write the variables $x_{i}\in\R^3$ as
$x_{i}=(x^\perp_{i},z_{i})$, where $ x^\perp_{i}\in\R^2$ and 
$z_{i}\in \R$ are respectively the components perpendicular and 
parallel to the field. 
Moreover, we write $(x_{1},\dots,x_{N})=(\xoperp,\underline z)$
with $\xoperp=(x^\perp_1,\dots,x^\perp_N)\in \R^{2N}$ and 
$\underline z=(z_{1},\dots,z_{N})\in\R^N$. 

In the lowest Landau band the part of (\ref{ha}) associated with the motion 
perpendicular to the field is exactly canceled by the spin contribution and 
only the
part corresponding to the motion along the field remains. Hence
\beq \Pi^0_N H_{B,Z,N}\Pi^0_N=\Pi^0_N H_{Z,N}\Pi^0_N\eeq
with
\beq H_{Z,N}=\sum_{i=1}^N \left(-\partial^2/\partial z_{i}^2-{Z\over 
\vert 
x_{i}\vert}\right)+\sum_{i<j}^N {1\over \vert 
x_{i}-x_{j}\vert}.\label{HZN}\eeq

The operator (\ref{HZN}) contains no derivatives perpendicular to the field
and hence the variables
$\xoperp$ can be regarded as parameters for a differential operator in the 
variables parallel to the field. For each $\xoperp$
such that $x^\perp_1,\dots,x^\perp_N$ are all different from 
zero, we consider the one-dimensional Hamiltonian 
\begin{equation}
H_{Z,N}(\xoperp)=\sum_{i=1}^N 
\left(-{\partial_{z_i}}^2-{Z\over \sqrt{ 
z_i^2+(x^\perp_i)^2}}\right)+\sum_{i<j}^N {1\over  
\sqrt{(z_i-z_j)^2+(x^\perp_i-x^\perp_j)^2}}\label{hznx}
\end{equation}
acting on $\bigotimes\limits^N L^2(\R)=L^2(\R^N)$. 
The expectation 
values of $H_{Z,N}$ can be written as
\beq 
\langle\Psi,H_{Z,N}\Psi\rangle= \int\langle 
\Psi(\xoperp,\cdot),H_{Z,N}(\xoperp)
\Psi(\xoperp,\cdot)\rangle_{L^2(\R^N)}\, d\xoperp.\label{exp}
\eeq

The next step is a scaling of the variables. In the lowest Landau 
level the characteristic length in the directions perpendicular to 
the field is $B^{-1/2}$. One can therefore expect that for the computation of
$E^{\rm conf}(B,Z,N)$, i.e., the infimum of (\ref{exp}) over (normalized) 
$\Psi\in {\cal H}^0_N$, the properties of $H_{Z,N}(\xoperp)$ for 
$|x^\perp_i|\sim B^{-1/2}$ 
are decisive. Anticipating this, it is natural to make a transformation of 
variables, $(\xoperp,\underline z)\to (B^{1/2}\underline 
x^\perp,L(B)\underline z)$ where the scale factor $L(B)$ in the direction of 
the field has still to be 
specified. The corresponding unitary operator on $L^2({\mathbb R}^N)$ is  
\beq 
U\Psi(\underline z)=L(B)^{1/2}\Psi(L(B)\underline z),
\eeq
and the Hamiltonian transforms in the following way:
\beq U^{-1}H_{Z,N}(\xoperp)U= L(B)^2 h^B_{Z,N}(B^{1/2}\xoperp)\label{uneq}\eeq
where
\beq 
h^B_{Z,N}(\yoperp)=\sum_{i=1}^{N}\left(-\partial_{z_{i}}^2-
ZV_{B,|y_{i}^\perp|}(z_{i})\right)+
\sum_{i<j}V_{B,|y_{i}^\perp-y_{j}^\perp|}(z_{i}-z_{j})\label{hbzny}\eeq
and the potential $V_{B,r}(z)$ is (for $r>0$) defined as
\beq V_{B,r}(z)={L(B)}^{-1}(B^{-1}L(B)^2r^2+z^2)^{-1/2}.\label{vbr}
\eeq

Let $E_{Z,N}(\xoperp)$ and $e^B_{Z,N}(\yoperp)$ denote the ground 
state energies of $H_{Z,N}(\xoperp)$ and $h^B_{Z,N}(\yoperp)$ respectively. In 
order to avoid discussions about the domains of the Hamiltonians, which in 
fact 
depend on whether some of the parameters $x^\perp_i$ (resp.\ $y^\perp_i$) 
coincide, 
we define the ground state energies in terms of quadratic forms in the same 
way 
as (\ref{ezn}):
\beqa E_{Z,N}(\xoperp)&=&\inf \{\langle\Psi,H_{Z,N}(\xoperp)\Psi\rangle\,:\, 
\Psi\in 
C^{\infty}_{0}(\R^{N}),\, \Vert\Psi\Vert_{2}=1\},\\
e^B_{Z,N}(\yoperp)&=&\inf \{\langle\Psi,h^B_{Z,N}(\yoperp)\Psi\rangle\,:\, 
\Psi\in 
C^{\infty}_{0}(\R^{N}),\, \Vert\Psi\Vert_{2}=1\}.
\eeqa
These energies are connected by the scaling relation
\beq 
E_{Z,N}(B^{-1/2}\yoperp)/L(B)^{2}=e^B_{Z,N}(\yoperp).\label{scal}\eeq

In the next section we show that with the choice $L(B)\sim \ln B$ the 
potential 
$V_{B,r}(z)$ converges for each $r>0$ in the sense of distributions to the 
delta 
function as $B\to 0$. This is the heuristic basis of Theorem \ref{1}. Since 
the 
convergence is not uniform in $r$, however, more is needed for a rigorous 
proof. 
In particular, one needs estimates on the $r$-dependence of the convergence 
$V_{B,r}(z)\to \delta(z)$. These estimates,  stated in Lemmas \ref{lm:delta1} 
and \ref{cl:delta2} in the next section, can be regarded as variants 
of Propositions 3.3 and 3.4 in \cite{LSY94} and the Appendix in \cite{JY}, 
adapted to the problem at hand. They are included here for completeness.

The upper bound on the energy, given in Section 3, is a straight-forward 
variational calculation. The lower bound is more subtle.
An important ingredient needed is the superharmonicity of the energy 
$E_{Z,N}(\xoperp)$ in the variables $x^\perp_i$. 
This result, established in Theorem \ref{thm:super}, generalizes a 
corresponding 
result 
(Proposition 2.3) in \cite{LSY94}. 
Superharmonicity implies that the lowest value of  $E_{Z,N}(B^{-1/2}\yoperp)$ 
for $|y^\perp_i|\geq \varepsilon$ with $\varepsilon>0$ is obtained at the 
boundary of the variable range, i.e., when either $|y^\perp_i|= \varepsilon$ 
or 
$|y^\perp_i|\to\infty$. 
Variables tending to infinity can be ignored, since  $V_{B,r}(z)\to 0$ for 
$r\to\infty$, so by this result one may in (\ref{exp}) restrict the attention 
to 
wave functions localized where $|x^\perp_i|\leq \hbox{(\rm const.)}B^{-1/2}$. 
On 
the other hand, the requirement that only wave functions in the lowest 
Landau band are taken into account in (\ref{exp}) plays the role of a 'hard 
core 
condition' that prevents collapse, since such wave functions cannot be 
concentrated on shorter scales than $O(B^{-1/2})$. This  statement is made 
precise in Lemma \ref{5.3}. 

The lower bound is obtained in Section 5 by combining Theorem \ref{thm:super}, 
Lemma \ref{5.3} and the convergence of the potentials $V_{B,r}$ discussed in 
Section 2. It is noteworthy that this lower bound 
holds also for bosonic statistics while the upper bound holds for fermionic 
statistics, so that altogether the convergence of $E(Z,N,B)/(\ln B)^2$ to 
$e(Z,N)$ is independent of the statistics.

In section 6 we discuss the delta-function model (\ref{dh}) and in 
particular prove Theorem \ref{2}. In the course of the proof we compare 
(\ref{dh}) with another model, whose ground state energy can be explicitly 
calculated. This model provides an upper bound 
for the ground state energy of (\ref{dh}) and has the same mean field limit. 
The Hamiltonian for this model is
\begin{equation}\label{symm}
\widetilde{h}_{Z,N}=\sum_{i=1}^{N}\left( p_{i}^{2}-\delta
(z_{i})\right) +\frac{1}{2Z}\sum_{i<j}\delta (|z_{i}|-|z_{j}|).
\end{equation} 
An interesting feature of this model is the fact that the maximal number 
$N_{\rm c}$ of electrons that a nucleus of charge $Z$ 
can bind is exactly the largest integer satisfying
\beq N_{\rm c}<2Z+1.\eeq
(This fact is unrelated to Lieb's upper bound \cite{L83} for the maximal 
negative ionization of atoms that does not apply to the Pauli Hamiltonian with 
a homogeneous magnetic field.)  A corresponding statement for the Hamiltonian 
(\ref{dh}) is not known, except 
in the mean field limit, cf.\ Theorem 1.2. In this connection 
it should be mentioned that an estimate of the form form $N_{\rm c}<2Z+1+({\rm 
const.})\,B^{1/2}$ has been derived in \cite{BR} for a Hamiltonian of a 
similar type as (\ref{hbzny}). 
\section{The high $B$ limit of the Coulomb interaction}

We define the scaling factor $L(B)$ in the potential (\ref{vbr}) as the 
solution 
of the equation
\beq B^{1/2}=L(B)\sinh[L(B)/2].
\eeq
Since $\int_{0}^1 (a^2+z^2)^{-1/2}dz=\sinh^{-1}(1/a)$, we have with this 
choice
\beq \int_{|z|\leq 
r}V_{B,r}(z)dz=1.\label{one}\eeq
for all $B$. Note also that
\beq L(B)=\ln B+O(\ln\ln B)\eeq
as $B\to \infty$.

Let $\psi\in H^1({\bf R})=\{\psi: 
\int|\psi|^2+\int|d\psi/dz|^2<\infty\}$. 
Every such $\psi$ is a continuous function on $\R$.
\begin{lm}[Delta approximation, part 1]\label{lm:delta1}
\beq \Big||\psi(0)|^2-\int V_{B,r}(z)|\psi(z)|^2dz\Big|\leq
L(B)^{-1}\left[\lambda r^{-1}+8{}\lambda^{1/4} T^{3/4} 
r^{1/2}\right]\label{deltapprox}\label{approx1}\eeq
with $\lambda=\int|\psi|^2$, $T= \int|d\psi/dz|^2$.        
\end{lm}
\begin{proof} It suffices to take $r=1$, for the general case follows 
by scaling $z\to rz$. Write the difference on the left side of 
(\ref{deltapprox}) as $A_{1}+A_{2}$ with
\beqa A_{1}&=&-\int_{|z|\geq 1}V_{B,1}(z)|\psi(z)|^2dz,\\
A_{2}&=&\int_{|z|\leq 
1}V_{B,1}(z)\left[|\psi(0)|^2-|\psi(z)|^2\right]dz.\eeqa
The missing term
\beq A_{3}=\left[1-\int_{|z|\leq 1}V_{B,1}(z) dz\right]|\psi(0)|^2\eeq
is zero because of (\ref{one}). Since $|V_{B,1}(z)|\leq L(B)^{-1}$ 
for $|z|\geq 1$, we have 
\beq |A_{1}|\leq \lambda L(B)^{-1}.\label{A1}\eeq
For $|z|\leq 1$ we have in any case
\beq |V_{B,1}(z)|\leq L(B)^{-1}|z|^{-1}.\eeq
Moreover,
\beqa \Big||\psi(z)|^2-|\psi(0)|^2\Big|&\leq&|\psi(z)-\psi(0)| 
\left[|\psi(z)|+|\psi(0)|\right]\nonumber\\
&\leq&  \Big|\int_{0}^z \frac{d\psi}{dz'} dz'\Big|\cdot 
2\left[\int_{-\infty}^\infty \frac{d|\psi(z')|^2}{dz'} 
dz'\right]^{1/2}\nonumber\\
& \leq&|z|^{1/2}T^{1/2}2{}\lambda^{1/4}T^{1/4}=
2{}\lambda^{1/4}T^{3/4}|z|^{1/2}.\eeqa
Hence
\beq |A_{2}|\leq {2{}}{L(B)}^{-1}\left(\int_{|z|\leq 
1}|z|^{-1/2}dz\right)\lambda^{1/4}T^{3/4}=
8{} L(B)^{-1}\lambda^{1/4}T^{3/4}.\eeq
Combining the estimates for $A_{1}$ and $A_{2}$ gives 
(\ref{deltapprox}).
\end{proof}

\begin{lm}[Delta approximation, part 2]\label{cl:delta2} Let $\Psi\in 
H^1(\R^2)$ and 
put
\beq \lambda =\int\int |\Psi(z,z')|^2dzdz',\qquad 
T=\int\int |\partial_{z'}\Psi(z,z')|^2dzdz'.\eeq
Then
\beqa\Big |\int|\psi(z,z)|^2 dz  -\int\int 
V_{B,r}(z-z')|\psi(z,z')|^2 
dzdz'\Big|\nonumber\\
\leq L(B)^{-1}[\lambda r^{-1}+8{}\lambda^{1/4} T^{3/4} 
r^{1/2}].\label{approx2}\eeqa
\end{lm}
\begin{proof} Put $\lambda(z)=\int |\Psi(z,z')|^2dz'$, 
$T(z)=\int |\partial_{z'}\Psi(z,z')|^2dz'$. By (\ref{deltapprox}) we 
have
\beqa \Big ||\Psi(z,z)|^2   -\int V_{B,r}(z-z')|\Psi(z,z')|^2 
dz'\Big|\nonumber\\
\leq L(B)^{-1}[\lambda(z) r^{-1}+8{}\lambda(z)^{1/4} T(z)^{3/4} 
r^{1/2}].
\eeqa
Integration over $z$, using the H\"older inequality to estimate $\int 
\lambda(z)^{1/4}T(z)^{3/4}dz$, gives (\ref{approx2}).
\end{proof}

\section{Upper Bound}

Let $\psi\in{\cal S}(\R^{2N})$ be a smooth and rapidly decreasing wave 
function in the lowest Landau level at field strength 1, and let 
$\phi\in{\cal C}^\infty_0(\R^N)$. If $\psi$ and $\phi$ are 
normalized, i.e., $\int_{\R^{2N}}|\psi|^2=\int_{\R^{N}}|\phi|^2=1$, then
\beq\Psi_{B}(\underline x^\perp,\underline z)=(BL(B))^{N/2}
\psi(B^{1/2}\underline x^\perp)\phi(L(B)\underline
z)\label{ans}\eeq
is a normalized wave function in the lowest Landau band at field 
strength $B$. Moreover, using (\ref{exp}) and (\ref{uneq}) we have 
\begin{eqnarray}
E(B,Z,N)&\leq& \langle\Psi_{B},H_{{Z,N}}\Psi_{B}\rangle\nonumber\\
&=&L(B)^2\int |\psi(\underline 
y^\perp)|^2\langle\phi,h^B_{Z,N}(\underline 
y^\perp)\phi\rangle d^{2N}\underline 
y^\perp\nonumber
\end{eqnarray}
where $h^B_{Z,N}(\underline 
y^\perp)$ is given by (\ref{hbzny}). Since $L(B)^2/(\ln B)^2\to 1$ as 
$B\to\infty$ 
and $\psi$ is normalized, one has for the upper bound in Theorem \ref{1} only 
to 
check that
\begin{eqnarray}
\int |\psi(\underline 
y^\perp)|^2 V_{B,|y^\perp_{i}|}(z_{i})|\phi(\underline z)|^2 
d^{2N}\underline 
y^\perp d^{N}\underline z\nonumber\\
\to \int \delta(z_{i}) |\phi(\underline z)|^2 d^{N}\underline z\nonumber
\end{eqnarray}
and
\begin{eqnarray}
\int |\psi(\underline 
y^\perp)|^2V_{B,|y^\perp_{i}-y^\perp_{j}|}(z_{i}-z_{j})|\phi(\underline 
z)|^2 d^{2N}\underline 
y^\perp d^{N}\underline z\nonumber\\
\to \int \delta(z_{i}-z_{j}) |\phi(\underline z)|^2 d^{N}\underline z\nonumber
\end{eqnarray}	
as $B\to\infty$. But this is taken care of by Lemmas \ref{lm:delta1} and 
\ref{cl:delta2}. (That $V_{B,r}(z)$ is not defined for $r=0$ is of no 
consequence here, because the error terms in (\ref{approx1}) and 
(\ref{approx2}) 
are integrable 
all the way to $r=0$.)
We therefore have
\begin{proposition}[Upper bound]
\beq\liminf_{B\to\infty}\frac{E(B,Z,N)}{(\ln B)^2}\leq e(Z,N).\eeq
\end{proposition}

{\it Remark.} It is clear that our upper bound holds for fermions, although 
$e(Z,N)$ is the bosonic ground state energy of (\ref{dh}). In fact, in the
ansatz (\ref{ans}) above we may choose $\psi$ to be antisymmetric and 
$\phi$ 
to be symmetric; then $\Psi_B$ is antisymmetric. Note also that for the 
Hamiltonian (\ref{dh}) the bosonic ground state  energy is the same as its 
ground 
state energy without symmetry restriction.
\section{Superharmonicity}
In this section we take a closer look at the dependence of the ground state 
energy $E_{Z,N}(\xoperp)$ of the Hamiltonian (\ref{hznx}) on the parameter 
$\xoperp$. We start with a simple estimate:
\begin{lm}[Simple bounds]\label{lm:bound}
The function $\xoperp\mapsto E_{Z,N}(\xoperp)$ 
satisfies the bounds 
\begin{equation}\label{eq:Ebounds}
  -\sum_{i=1}^N Z^2\left(1+\left[\sinh^{-1}((Z|x_i^\perp|)^{-1})
    \right]^2\right)\leq E_{Z,N}(\xoperp)\leq0
\end{equation}
on the set 
\begin{equation}\label{eq:aset}
  {\cal A}=\{\xoperp\in\R^{2N}\ :\ x^\perp_i\ne0,\text{ for all
  }i=1,\ldots,N\}.
\end{equation}
\end{lm}
\begin{proof}The non-positivity of $E$ is straightforward from the definition 
by an appropriate choice of $\Psi$. Note that this also holds in the
case where some of the $x^\perp_i$ variables coincide.
The lower bound on $E_{Z,N}(\xoperp)$ follows from Lemma~2.1 in 
\cite{LSY94} together with the operator inequality
$$
 H_{Z,N}(\xoperp)\geq \sum_{i=1}^N 
\left(-\partial_{z_i}^2-{Z\over \sqrt{ 
z_i^2+(x^\perp_i)^2}}\right)
$$
which is obtained by ignoring the positive two-body interactions.
\end{proof}

Next we turn to the superharmonicity properties of $E_{Z,N}(\xoperp)$.
We shall need the following general result. 
\begin{lm}[Inherited superharmonicity]
Let $U$ be an open set in $\R^d$ and assume that
$f:U\times\R\to(-\infty,\infty]$ is a superharmonic 
function with the property that 
$$
b=\min\{\liminf_{t\to\infty}f(x,t), 
\liminf_{t\to-\infty}f(x,t)\}
$$ 
is independent of $x$ for all 
$x\in U$. 
Then $g(x)=\inf_tf(x,t)$ is a superharmonic function 
on $U$.
\end{lm}
\begin{proof}
We shall prove this by showing that $\Delta g\leq0$ as 
a distribution. We shall use that $f$ is a 
lower semicontinuous function satisfying the mean 
value inequality
$$
        \int_{|(x,t)-(y,s)|\leq r}f(y,s) dyds\leq 
        f(x,t)c_{d+1}r^{d+1},
$$ 
for all $(x,t)\in U\times \R$ if $r>0$ is small enough,
where $c_{d+1}$ is the volume of the unit ball in $\R^{d+1}$.

For $x\in U$ it follows from the 
lower semicontinuity of $f$ that 
we have either $g(x)=b$ or there exists $t\in\R$ such that 
$g(x)=f(x,t)$. In the first case we obviously have 
\begin{eqnarray}\label{eq:g}
        c_{d+1}r^{d+1}g(x)
        \geq 2\int_{|x-y|\leq r}g(y)\sqrt{r^2-(x-y)^2}dy
\end{eqnarray}
since $g(y)\leq b$ for all $y$.
If $g(x)<b$ we also conclude the above inequality since
\begin{eqnarray*}
         g(x)c_{d+1}r^{d+1}&=&f(x,t)c_{d+1}r^{d+1}
        \geq\int_{|(x,t)-(y,s)|\leq r}f(y,s) dyds\\
        &\geq& \int_{|(x,t)-(y,s)|\leq r}g(y) dyds
        =2\int_{|x-y|\leq r}g(y)\sqrt{r^2-(x-y)^2}dy.
\end{eqnarray*}
Note now that for any $\phi\in C^\infty_0(U)$
we have for any $x\in U$ that 
$$
        \lim_{r\to0}r^{-(d+3)}
        \int_{|x-y|\leq r}[\phi(y)-\phi(x)]\sqrt{r^2-(x-y)^2}dy
        =C \Delta \phi(x)
$$
for some constant $C>0$
and in fact this limit holds in the topology of 
$C^\infty_0(U)$.
Thus if $\phi\geq0$
we have 
$$
        \int_U g(x)\Delta\phi(x)dx
        =C^{-1}\lim_{r\to0}r^{-(d+3)}
        \int_{|x-y|\leq r} g(x)
        (\phi(y)-\phi(x))\sqrt{r^2-(x-y)^2}dydx\leq0
$$
by the inequality (\ref{eq:g}). Hence $\Delta g\leq 0$.
\end{proof}

\begin{thm}[Superharmonicity of the energy]\label{thm:super}
On the set ${\cal A}$ defined in (\ref{eq:aset}) the function $\xoperp\mapsto 
E_{Z,N}(\xoperp)$ is 
superharmonic in each of the variables 
$x^\perp_i$, $i=1,\ldots,N$
independently.
\end{thm}
\begin{proof}

We follow closely the proof of
Prop.~2.3 in \cite{LSY94}, which stated the superharmonicity of 
the ground state energy of a one-body operator which 
can be considered as a mean field approximation of $H_{Z,N}(\xoperp)$.

It is clearly enough to prove that $E_{Z,N}(\xoperp)$ is 
superharmonic in $x^\perp_1$ (on the region $x^\perp_1\ne0$)
for $x^\perp_2,\ldots,
x^\perp_N$ fixed. 
We shall prove this by showing that $x^\perp_1\mapsto 
E_{Z,N}(\xoperp)$ 
satisfies the mean value inequality around any given point 
$x^\perp_{1,0}$. Let $\xoperp_0=(x^\perp_{1,0},x^\perp_2,\ldots,
x^\perp_N)$.
Choose a sequence of $L^2$ normalized functions 
$\Psi_n\in C^\infty_0(\R^N)$ such that
$\langle\Psi_n,H_{Z,N}(\xoperp_0)\Psi_n\rangle\to E_{Z,N}(\xoperp_0)$ as 
$n\to\infty$.

For $w\in \R$ denote by $\Psi^{(w)}_n$ the function 
$$
  \Psi^{(w)}_n(z_1,\ldots,z_N)=\Psi_n(z_1-w,z_2,\ldots,z_N).
$$
We clearly have 
$$
  \inf_{w\in\R}
  \langle\Psi^{(w)}_n,H_{Z,N}(\xoperp_0)\Psi^{(w)}_n\rangle\to 
E_{Z,N}(\xoperp_0)
  \quad\text{as }n\to\infty.
$$

If $x^\perp_1$ is close to $x^\perp_{1,0}$ we shall use 
$\Psi^{(w)}_{n}$ as a trial function for $H(\xoperp)$. 
We then obtain 
$$
   E_{Z,N}(\xoperp)\leq \liminf_n\inf_{w\in\R}
   \langle\Psi^{(w)}_n,H_{Z,N}(\xoperp)\Psi^{(w)}_n\rangle.
$$
Hence 
\begin{equation}\label{eq:wnbound}
  E_{Z,N}(\xoperp)-E_{Z,N}(\xoperp_0)\leq \liminf_n\left[
  \inf_{w\in\R}\langle\Psi^{(w)}_n,H_{Z,N}(\xoperp)\Psi^{(w)}_n\rangle
-\inf_{v\in\R}\langle\Psi^{(v)}_n,H_{Z,N}(\xoperp_0)\Psi^{(v)}_n\rangle\right]
.
\end{equation}
The potential appearing in $H_{Z,N}(\underline{x}^\perp)$,
 i.e.,
$$
    W_{Z,N,\xoperp}(z_1,\ldots,z_N)=-\sum_{i=1}^N{Z\over \sqrt{ 
z_i^2+(x^\perp_i)^2}}+\sum_{i<j}^N {1\over  
\sqrt{(z_i-z_j)^2+(x^\perp_i-x^\perp_j)^2}}.
$$  
is a superharmonic
function of $(z_1,x^\perp_1)\in \R^3\setminus\{0\}$.
Writing
$$
  \langle\Psi^{(w)}_n,W_{Z,N,\xoperp}\Psi^{(w)}_n\rangle=
  \int W_{Z,N,\xoperp}(z_1+w,z_2,\ldots,z_N)|\Psi_n(z_1,\ldots,z_N)|^2
  dz_1\cdots dz_N
$$
we see that $\langle\Psi^{(w)}_n,W_{Z,N,\xoperp}\Psi^{(w)}_n\rangle$ 
is superharmonic in $(w,x^\perp_1)$ away from the line $x^\perp_1=0$.
Since  $\langle\Psi^{(w)}_n,\partial_{z_i}^2\Psi^{(w)}_n\rangle$
is independent of $w$ and $x^\perp_1$ for all $i=1,\ldots,N$ we have 
that $\langle\Psi^{(w)}_n,H_{Z,N}(\xoperp)\Psi^{(w)}_n\rangle$
is superharmonic in $(w,x^\perp_1)$ away from the line $x^\perp_1=0$.

Moreover, we also have that the two limits
$$
\liminf_{w\to\pm\infty}\langle\Psi^{(w)}_n,H_{Z,N}(\xoperp)\Psi^{(w)}_n\rangle
$$
are independent of $x^\perp_1$. This is true simply because the 
contribution from the terms in the Hamiltonian depending on
$x^\perp_1$
tend to zero as $w\to\pm\infty$.
We may therefore apply the above lemma to the function
$f(w,x^\perp_1)=\langle\Psi^{(w)}_n,H_{Z,N}(\xoperp)\Psi^{(w)}_n\rangle$.
We conclude that
the function
$$
  x^\perp_1\mapsto
  \inf_{w\in\R}\langle\Psi^{(w)}_n,H_{Z,N}(\xoperp)\Psi^{(w)}_n\rangle
$$
is superharmonic for $x^\perp_1\ne0$. Moreover by the inequality
(\ref{eq:Ebounds}) this 
function is bounded below if $|x^\perp_1|$ is bounded away from $0$.

Now using Fatou's Lemma we see from (\ref{eq:wnbound}) that the 
average of 
$E_{Z,N}(\xoperp)-E_N(\xoperp_0)$ over 
the set $\{x^\perp_1\ :\ |x^\perp_1-x^\perp_{1,0}|<r\}$ is
non-positive for all $r>0$ small enough.
\end{proof}

\section{Lower Bound}
The first lemma in this section concerns the ground state energy 
$e^B_{Z,N}(\yoperp)$ of $h^B_{Z,N}(\yoperp)$ and does not use 
superharmonicity.

\begin{lm}[Lower bound on $e^B_{Z,N}(\yoperp)$] \label{lm:eblower} Let ${\cal 
K}$ 
be a compact subset of 
the set 
${\cal A}$
given in (\ref{eq:aset}). Then
        \beq \liminf_{B\to\infty}\inf_{\yoperp\in{\cal K}}
        e^B_{Z,N}(\yoperp)\geq e(Z,N).\eeq 
\end{lm}

\begin{proof} To avoid problems at points $\yoperp$ with 
$y_{i}^\perp-y_{j}^\perp=0$ for some $i,j$, we replace the repulsive 
potential $V_{B,|y_{i}^\perp-y_{j}^\perp|}(z_{i}-z_{j})$ by the 
smaller potential $V_{B,|y_{i}^\perp-y_{j}^\perp|+1}(z_{i}-z_{j})$. 
 We denote the 
corresponding Hamiltonian by $\tilde h^B_{Z,N}(\yoperp)$ and its 
ground 
state energy by $\tilde e^B_{Z,N}(\yoperp)$.  It is obvious that
$e^B_{Z,N}(\yoperp)\geq \tilde e^B_{Z,N}(\yoperp)$, so a lower bound 
on 
$\tilde 
e^B_{Z,N}(\yoperp)$ gives a lower bound on $e^B_{Z,N}(\yoperp)$.

Let $\Psi$ be a normalized, symmetric wavefunction in $C^\infty_0({\R}^N)$. 
Since $\langle \Psi,h_{Z,N} 
\Psi\rangle\geq e(Z,N)$ we have to estimate the matrix 
elements of the difference $\tilde h^B_{Z,N}(\yoperp)-h_{Z,N}$. Using 
Lemmas \ref{lm:delta1} and \ref{cl:delta2}, together with the
H\"older inequality for the integration over  $z_2,\dots,z_N$ and 
$z_3,\dots,z_N$ respectively, we obtain
\beqa
\left |\langle \Psi, \tilde h^B_{Z,N}
\Psi\rangle-\langle \Psi,h_{Z,N} 
\Psi\rangle\right|\leq L(B)^{-1}(ZN+N(N-1))\nonumber \\ \times
\left[r_{\rm min}^{-1}+8{} T_\Psi^{3/4} (2r_{\rm max}+1)^{1/2} 
\right]\label{error}
\eeqa
where $r_{\rm min}$ and $r_{\rm max}$ are respectively the minimum and
the maximum value of $|y^\perp_i|$, $i=1,\dots,N$,  with $\yoperp\in{\cal K}$, 
and 
\beq T_\Psi=N\int|\partial_{z}
\Psi(z,z_2,\dots,z_N)|^2dz dz_2\cdots dz_N\eeq
is the kinetic energy of $\Psi$. Now if $\Psi^{B}_{\yoperp,n}$, 
$n=1,2,\dots$ is a minimizing sequence of 
normalized wave functions for $\tilde h^B_{Z,N}(\yoperp)$, then we may assume 
that the
corresponding kinetic energy is uniformly bounded in $n$,
$B$ and $\yoperp\in{\cal K}$.
In fact, we may assume that $\langle\Psi^{B}_{\yoperp,n},
\tilde h^B_{Z,N}(\yoperp)\Psi^{B}_{\yoperp,n}\rangle$
is a bounded sequence. If we use the bound
from Lemma 2.1 in [LSYa], we obtain
\beq
\langle\Psi^{B}_{\yoperp,n},
\tilde h^B_{Z,N}(\yoperp)\Psi^{B}_{\yoperp,n}\rangle
	\geq \frac{T_n}{2}-\frac{1}{2}\left(\frac{2Z}{L(B)}\right)^2
\left(1+\left[\sinh^{-1}\{(2Z)^{-1}B^{1/2}\}\right]^2\right),
\eeq
where we have saved half of the
the kinetic energy $T_n$ of $\Psi^{B}_{\yoperp,n}$.
For large $B$, $L(B)^{-1}\left[\sinh^{-1}\{(2Z)^{-1}B^{1/2}\}\right]$
is bounded and hence we see that $T_n$ is bounded. The error term 
(\ref{error}) 
with $\Psi=\Psi^{B}_{\yoperp,n}$ thus tends to zero as $B\to\infty$,
uniformly in $n$, and the lemma is established.
\end{proof}

\begin{lm}[Uniform bounds on $E_{Z,N}(\xoperp)$] \label{5.2}Let 
$\varepsilon>0$.
Consider the set 
\beq
      {\cal C}^{B,\varepsilon}=\{\xoperp\ :\ \varepsilon B^{-1/2}\leq\
        |x^\perp_i|,
        \text{ for all }i=1,\ldots,N\}.\label{cbe}
\eeq
Then 
\beq
  \liminf_{B\to\infty}(\ln B)^{-2}\inf\{E_{Z,N}(\xoperp)\ : \
        \xoperp\in {\cal C}^{B,\varepsilon}\}\geq e(Z,N).\label{unif}
\eeq
where $e(Z,N)$ as before denotes the 1-dimensional delta function atom 
energy.
\end{lm}
\begin{proof}
Define the sets 
$$
        {\cal C}^{B,\varepsilon}_n=\{\xoperp\ :\ \varepsilon 
B^{-1/2}\leq\
        |x^\perp_i|\leq n,
        \text{ for all }i=1,\ldots,N\}.
$$
Since ${\cal C}^{B,\varepsilon}_n$ is compact and $ E_{Z,N}$ is lower 
semicontinuous (being superharmonic, in fact, superharmonic in each 
variable) we may find 
$\xoperp_n\in {\cal C}^{B,\varepsilon}_n$ 
such that 
$$
        E_{Z,N}(\xoperp_n)=\min\{E_{Z,N}(\xoperp)\ : \
        \xoperp\in {\cal C}^{B,\varepsilon}_n\}.
$$
Clearly, 
$$
       \lim_{n\to\infty} E_{Z,N}(\xoperp_n)\to
       \inf\{E_{Z,N}(\xoperp)\ : \
        \xoperp\in {\cal C}^{B,\varepsilon}\}.
$$
By the superharmonicity of $E_{Z,N}(\xoperp)$ in each variable
$x^\perp_i$ we know  that each coordinate 
$x^\perp_{i,n}$ of the point $\xoperp_n$ satisfies either 
$|x^\perp_{i,n}|=\varepsilon B^{-1/2}$ or
$|x^\perp_{i,n}|=n$. Moreover, since $E_{Z,N}(\xoperp)$ is invariant under 
permutations of the 
coordinates of $\xoperp$ we may assume that $|x^\perp_{1,n}|\leq 
|x^\perp_{2,n}|\leq \dots\leq |x^\perp_{N,n}|$ for all $n$.
By possibly going to a subsequence 
we may assume that there exists an integer $0\leq K\leq N$ such that 
for $n$ large enough 
$$
 |x^\perp_{i,n}|=\left\{
 \begin{array}{ll}\varepsilon B^{-1/2},&\text{ for }i=1,\ldots,K\\
   n,&\text{ for }i>K\\
 \end{array}\right. .
$$
Moreover, we may assume that $ x^\perp_{i,n}$ converges as 
$n\to\infty$ for $i=1,\ldots,K$. 

Since we may ignore the variables $x^\perp_{i,n}$, $i=K+1,\ldots,N$, 
which tend to infinity we have 
$$
   \lim_{n\to\infty}E_{Z,N}(\xoperp_n)/E_{Z,K}(x^\perp_{1,n},\ldots,
   x^\perp_{K,n})=1
$$
Since $E_{Z,K}(\xoperp)$ is lower 
semicontinuous we conclude 
that there exists a point $(x^\perp_{1,\infty},\ldots,
x^\perp_{K,\infty})\in\R^{2K}$ with 
$|x^\perp_{i,\infty}|= \varepsilon B^{-1/2}$ for all $i=1,\ldots,K$ 
such that 
$$
 \inf\{E_{Z,N}(\xoperp)\ : \
        \xoperp\in {\cal C}^{B,\varepsilon}\}= 
E_{Z,K}(x^\perp_{1,\infty},\ldots,
         x^\perp_{K,\infty}).
$$

By Lemma~\ref{lm:eblower} we have that
$$
       \liminf_{B\to\infty}
       \inf\{L(B)^{-2}E_{Z,K}(B^{-1/2}y^\perp_{1},\ldots,
         B^{-1/2}y^\perp_K)\ : \ |y^\perp_{i}|= \varepsilon,
           \text{ for all }i\}\geq e(Z,K).
$$
Since $K\leq N$ and hence $e(Z,K)\geq e(Z,N)$ we have proved 
the lemma.
\end{proof}

\begin{lm}[Wave functions in the lowest Landau band]\label{5.3}
If  
$\Psi\in{\cal H}_N^0$ belongs to the lowest Landau band at field strength $B$, 
then 
$\int|\Psi(\xoperp,\underline{z})|^2
d\underline{z}$ is a bounded function of $\xoperp$ (possibly after a 
modification on a null set) and for all $1\leq n\leq N$
\beq \label{lbound}
\sup_{x_1^\perp,\dots,x_n^\perp}\Big|\int\int|
\Psi(\xoperp,\underline{z})|^2
d\underline{z}dx_{n+
1}^\perp\cdots dx_{N}^\perp\Big |\leq \frac {B^n}{(2\pi)^n} \Vert 
\Psi\Vert^2\eeq
\end{lm}
\begin{proof} The projector $\Pi^0_N$ on the lowest Landau band is the $N$-th 
tensorial
power of the projector $\Pi^0$ that operates on $L^2({\mathbb R}^3;{\mathbb 
C}^2)$ and is given by the integral kernel
\beq\label{intkern}\Pi^0(x,x')=\Pi_\perp^0(x^\perp,{x'}^\perp)\delta(z-z')
P^{\downarrow},
\eeq
where
\beq\label{intkern2}
\Pi_\perp^0(x^\perp,{x'}^\perp)=\frac B{2\pi}\exp\left\{\mfr{\rm 
i}/2(x^\perp\times 
{x'}^\perp)\cdot{\bf B}-\mfr 1/4(x^\perp-{x'}^\perp)^2B\right\}
\eeq
and $P^{\downarrow}$ is the the projector on vectors in ${\mathbb C}^2$ with 
spin component $-1/2$. The kernel $\Pi_\perp^0(x^\perp,{x'}^\perp)$ is a 
continuous function with
\beq\label{proj}\int 
\Pi^0(x^\perp,u^\perp)\Pi^0(u^\perp,y^\perp)du^\perp=\Pi^0(x^\perp,y^\perp)
\eeq 
and
\beq\label{bd}
\Pi^0(x^\perp,x^\perp)=\frac B{2\pi}
\eeq
for all $x^\perp$.
A wave function in the lowest Landau band has the representation
$\Psi=\Pi^0_N\Psi$. After writing $\Pi^0_N$ as an integral operator 
(\ref{lbound}) follows from the Cauchy-Schwarz inequality, using (\ref{proj}) 
and (\ref{bd}).
\end{proof}

\begin{proposition}[Lower bound]
\beq\liminf_{B\to\infty}\frac{E(B,Z,N)}{(\ln B)^2}\geq e(Z,N).\eeq
\end{proposition}
\begin{proof} For fixed $B$ let $\Psi$ be a normalized wave function 
in the
  lowest Landau band. By (\ref{exp}) we have
\beq\langle\Psi,H_{Z,N}\Psi\rangle\geq\int E_{Z,N}(\xoperp)
\left(\int|\Psi(\xoperp,\underline
  z)|^2d\underline z\right)d\xoperp. 
\eeq
We split the integral over $\xoperp$ into an integral over ${\cal
  C}^{B,\varepsilon}$ (defined in (\ref{cbe})) and its complement in 
$\R^{2N}$. By Lemma
\ref{5.2} 
we have only to
consider the latter. Using the estimate (\ref{eq:Ebounds}) the task is to 
bound 
terms
of the form
\beq\int_{|x^\perp_i|\leq \varepsilon B^{-1/2}} 
(1+[\sinh^{-1}(Z|x_j^\perp|^{-1})]^2)\left(\int|\Psi(\xoperp,\underline
  z)|^2d\underline z\right)d\xoperp 
\eeq
from above. If $i=j$ we carry out the integration over all $x_k^\perp$
with $k\neq i$ and use Lemma \ref{5.3} for the remaining variable
$x^\perp_i$. For small $r$, $|\sinh^{-1} r^{-1}|\leq
{\rm(const.)}|\ln r|$
and the term can be estimated by
\beq {\rm(const.)}\int_{|x^\perp|\leq \varepsilon B^{-1/2}}(\ln
|x^\perp|)^2 B dx^\perp\leq {\rm(const.)}\varepsilon^2 (\ln B)^2.
\eeq
For $i\neq j$ we split the integration over $x^\perp_j$ into two
parts, namely $|x^\perp_j|\leq B^{-1/2}$ and $|x^\perp_j|\geq
B^{-1/2}$.
For the first part we obtain the following bound, after transforming variables 
and using Lemma 5.3, this time for $n=2$,
\beq {\rm(const.)}\varepsilon^2\int_{|y^\perp_i|\leq 1,|y^\perp_j|\leq 1
  }(\ln B^{-1/2}
|y_j^\perp|)^2 dy^\perp_i dy^\perp_j\leq {\rm(const.)}\varepsilon^2
(\ln B)^2.
\eeq
For the integral over   $|x^\perp_j|\geq B^{-1/2}$ we
estimate $|\sinh^{-1}( Z|x^\perp_j|^{-1})|^2$ by its  maximum
value,
$\leq {\rm(const.)}(\ln B)^2$ and obtain
for this part of the integral the upper bound
\beqa
	\int_{|x^\perp_i|<\varepsilon B^{-1/2}}
	(1+{\rm(const.)}(\ln B)^2)\left(\int|\Psi(\xoperp,\underline
  z)|^2d\underline z\right)d\xoperp\nonumber\\
	\leq \int_{|x^\perp_i|<\varepsilon B^{-1/2}}
	(1+{\rm(const.)}(\ln B)^2)Bd x^\perp_i\leq
	\varepsilon^2(1+c(\ln B)^2),
\eeqa
where we have used Lemma~\ref{5.3} again.
We see that (55) is bounded
above by ${\rm(const.)}(\varepsilon\ln B)^2$, for $B$ large enough.
Since $\varepsilon>0$ is arbitrary this completes the
proof.
\end{proof}

\section{The one-dimensional delta-function model}

We now want to study the the delta-function Hamiltonian 
(\ref{dh}), in particular its mean field limit, 
$N\to \infty $, $
Z\to \infty $, with $ \lambda =N/Z$ fixed.

For this it is convenient to make a scale transformation $z \rightarrow 
z/Z$, which implies a unitary equivalence 
\begin{equation}
h_{Z,N} \cong Z^2 \widehat{h}_{Z,N}
\end{equation}
with 
\begin{equation}
\widehat{h}_{Z,N} = \sum_{i=1}^N \left( p_i^2 - \delta(z_i)\right) + 
\frac{1%
}{Z} \sum_{i<j} \delta (z_i - z_j).\label{hhat}
\end{equation}

We denote its ground state energy (again in the sense of quadratic forms) 
by 
$\widehat{e}(Z,N)$. The formal mean field theory of this system is identical
to the so called hyper-strong theory discussed in \cite{LSY94}, Section 3.
The energy of a (one dimensional) electron density $Z\rho $ in this theory
is $Z\mathcal{E}^{\mathrm{HS}}[\rho ]$ with

\begin{equation}  \label{equation:hsf}
\mathcal{E}^{\mathrm{HS}}[\rho ] = \int_{\mathord{\mathbb R} }\left( 
\frac{d%
}{dz}\sqrt{\rho (z)}\right) ^{2}-\rho (0)+ \int \rho ^{2}dz
\end{equation}
The infimum over densities with fixed normalization $\int \rho = \lambda $
leads to the hyper-strong energy $E^{\mathrm{HS}}(\lambda )$ given by the
right side of (\ref{hs}).

We shall now establish this mean field limit rigorously and prove Theorem 1.2.

\subsection{A comparison model}

An upper bound to the Hamiltonian (\ref{dh}) can be obtained from 
another model whose ground state can be computed explicitly.  The 
corresponding Hamiltonian is completely symmetric with regard to each 
single reflection $z_{i}\rightarrow -z_{i}$, and the electronic 
repulsions are equally distributed between the sites $z_{i}$ and 
$-z_{i}$:
 
\begin{equation}
\widetilde{h}_{Z,N}=\sum_{i=1}^{N}\left( p_{i}^{2}-\delta (z_{i})\right) 
+%
\frac{1}{2Z}\sum_{i<j}\left[ \delta (z_{i}-z_{j})+\delta 
(z_{i}+z_{j})\right].\label{newh}
\end{equation}
Its ground state energy is denoted by $\widetilde{e}(Z,N)$.\bigskip 

The replacement of $1/Z$ by $1/(2Z)$ is important, because it compensates to 
a certain extent the doubling of the interaction sites. In particular it 
leads to the \textit{same} formal mean
field theory as (\ref{hhat}) for symmetric electron densities $\rho $. This 
observation will be substantiated by the mathematical treatment in the sequel.

The model (\ref{newh}) was used in \cite{WS70} for $N=2$ as a starting 
point for a perturbational calculation.  It was also considered in 
\cite{Ro71} (for $N=2$) as an upper bound to the model (\ref{dh}), but 
with the coupling $1/Z$ instead of $1/(2Z)$.  The present 
considerations and extensions to $N\geq 3$ appear to be new.\bigskip
 
The ground state wave function $\widetilde{\psi }$ , if it exists, is 
completely symmetric under permutations of $\{z_{1}...z_{N}\}$ 
and reflections $z_{i}\rightarrow -z_{i}$.
Such a highly symmetric function $\widetilde{\psi }$ is determined by its 
restriction 
to the cone
\begin{equation}
\label{cone} 
\mathcal{M} = \{\underline{z}:0\leq z_{1}\leq z_{2}\leq \ldots \leq z_{N}\}  
\end{equation}
In $\mathcal{M}$ we make the ansatz
\begin{equation}
\label{grs} 
\widetilde{\psi }(z_{1}\ldots z_{N})=c\prod_{i=1}^{N}
e^{-\kappa _{i}z_{i}} , 
\end{equation} 
with $c$ a normalization constant
and let $\widetilde{h}_{Z,N}$ act on the symmetrically extended 
$\widetilde{\psi }$. 

The delta-function interactions dictate the jumps in the partial 
logarithmic derivatives of $\widetilde{\psi }$ at the boundary of
$\mathcal{M}$ 
and we find
\[
\kappa _{1}=\frac{1}{2},\qquad \kappa _{i}-\kappa _{i-1}=-\frac{1}{4Z},
\]
which implies 
\begin{equation}
\kappa _{n}=\frac{1}{2}-\frac{n-1}{4Z}.  \label{kappa}
\end{equation}
The function (\ref{grs}) is square integrable 
if and only if all $\kappa_{n}$ are strictly positive, which is 
equivalent to
\begin{equation}
N<2Z+1.  \label{bindcond}
\end{equation}

The corresponding eigenfunction  $\widetilde{\psi }$, is everywhere 
positive, and it is easy to see that it is, indeed, a ground state 
for (\ref{newh}): 
Define the operators $A_{n}$ on $L^2(\R^N)$ by
\begin{equation}
A_{n}=\partial_{z_n}-\partial_{z_{n}}(\ln\widetilde\psi)\,,
\end{equation}
with obvious domains of definition. 
Denoting by $\widetilde{e}(\widetilde{\psi})$ the eigenvalue of 
$\widetilde{h}_{Z,N}$ corresponding to $\widetilde{\psi } $  
we can write the quadratic form 
$\widetilde{h}_{Z,N} $ as 
\begin{equation}\label{astara} 
\widetilde{h}_{Z,N}=\sum_{i=1}^{N} A_{n}^{\ast }A_{n} +%
\widetilde{e}(\widetilde{\psi}).
\end{equation} 
The equation 
\begin{equation}
\langle \psi ,\widetilde{h}_{Z,N}\psi \rangle =
\sum_{i=1}^{N} \parallel A_{n} \psi \parallel^{2} +\, 
\widetilde{e}(\widetilde{\psi}) \parallel \psi \parallel^{2} \ \geq %
\ \widetilde{e}(\widetilde{\psi}) \parallel \psi \parallel^{2}, 
\end{equation}
which holds for each $ \psi $ in the form domain of 
$\widetilde{h}_{Z,N}$, shows that 
$\widetilde{e}(Z,N) = \widetilde{e}(\widetilde{\psi})$.

If $N \geq 2Z+1$, the simple inequality 
$\widetilde{e}(Z,N) \leq \widetilde{e}(Z,N-1)$ is sufficient 
for our purposes. To prove it, one may use 
trial-wave-functions of the form 
$\psi (z_{1}\ldots z_{N-1})\varepsilon \varphi (\varepsilon ^2 z_{N})$ 
whith a smooth $\varphi $, and take $\varepsilon$ to zero. 
This inequality for the energies can be iterated to
\begin{equation}
\widetilde{e}(Z,N) \leq \widetilde{e}(Z,N_{\rm o}),
\end{equation}
where $N_{\rm o}$ is the largest integer satisfying (\ref{bindcond}).

For $N<2Z+1$ the \textit{ground state energy} is the 
eigenvalue corresponding to (\ref{grs}):
\begin{eqnarray}
\widetilde{e}(Z,N) &=& \widetilde{e}(\widetilde{\psi}) =%
-\sum_{n=1}^{N}\kappa _{n}^{2}=  \nonumber\\
&=&-\frac{1}{4}\left\{ N\left( 1-\frac{\lambda }{2}+\frac{\lambda 
^{2}}{12}%
\right) +\left( \frac{\lambda }{2}-\frac{\lambda ^{2}}{8}\right) +\frac{%
\lambda ^{2}}{24N}\right\}. \label{gse}
\end{eqnarray}
If $N\geq N_{\rm o}$ we may use (73), i.e., $\widetilde{e}(Z,N)$ is bounded 
from above by (\ref{gse}) with 
$\lambda$ replaced by $\lambda_{\rm o}=N_{\rm o}/Z$. 
Dividing by $Z$ and keeping $\lambda $ fixed, the leading term for 
$N\to\infty$ is identical
to $E^{\mathrm{HS}}(\lambda)$. 
By the next proposition this is sufficient for 
the upper bound in Theorem 1.2. But one can in fact show that
$\widetilde{e}(Z,N_{\rm o})$ is {\it equal} to $\widetilde{e}(Z,N)$ for 
$N\geq 2Z+1$ and not only an upper bound to it.  Hence $N_{\rm o}$ is 
equal to $N_c$, the maximal number of electrons that can be bound in the model 
(\ref{newh}). We give the proof of this result in the appendix.

\begin{proposition}
[The comparison model gives upper bounds] The ground state  energy  
of the symmetrized model $\widetilde{h}_{Z,N}$ is 
an
upper bound to the ground state energy of
$\widehat{h}%
_{Z,N}$: 
\begin{equation}
\widehat{e}(Z,N)\leq \widetilde{e}(Z,N).  \label{upperbound}
\end{equation}
This inequality is strict, if $N\geq 2$ and $N<2Z+1$.
\end{proposition}

\begin{proof}
The Hamiltonian $\widetilde{h}_{Z,N}$ is the symmetrization of 
$\widehat{h}%
_{Z,N}$ with respect to the group $\mathcal{R}$ with $2^N$ elements, 
generated by the 
reflections 
$z_{i}\rightarrow -z_{i}$, $i=1,\dots,N$. For $R\in \mathcal{R}$ let $U_{R}$
denote the corresponding unitary operators on ${L}^{2}(\R^{N})$ .
Then 
\[
\frac{1}{2^{N}}\sum_{R\in \mathcal{R}}\langle U_{R}\psi ,\widehat{h}%
_{Z,N}U_{R}\psi \rangle =\langle \psi ,\widetilde{h}_{Z,N}\psi \rangle 
\]
for any $\psi $, so 
\[
\widehat{e}(Z,N)\leq \widetilde{e}(Z,N). 
\]

If $N < 2Z+1$ we may take the square integrable ground state wave function  of 
$\widetilde{h}_{Z,N}$, given by (\ref{grs}),
as a test state for $\widehat{h}_{Z,N}$. It satisfies $U_{R}\widetilde{\psi 
}
=\widetilde{\psi }$ for all $R$, so 
\[
\langle \widetilde{\psi },\widehat{h}_{Z,N}\widetilde{\psi }\rangle =
\widetilde{e}(Z,N).
\]

But $\widetilde{\psi }$ is not an eigenfunction of $\widehat{h}_{Z,N}$  if 
$N\geq 2$, so $\widehat{e}(Z,N)$ is strictly below $\widetilde{e}(Z,N)$.
\end{proof}

Combining the last proposition with Eq.\ (\ref{gse}), recalling that 
$e(Z,N)=Z^2\widehat e(Z,N)$, we obtain
\begin{proposition}[Upper bound in the mean field limit] If $N,Z\to\infty$ 
with $\lambda=N/Z$ fixed, then
\beq\limsup e(Z,N)/Z^3\leq E^{\rm HS}(\lambda)\eeq
where $E^{\rm HS}(\lambda)$ is given by (\ref{hs}). 

\end{proposition}

\subsection{Lower bounds to the delta-function Hamiltonian}

An elegant way to obtain lower bounds for Hamiltonians with 
repulsive pair interactions is
the use of positive definite functions. This was probably 
done for the first time
in \cite{HLT75}. In this method, the positive definite 
functions have to be
finite at the origin, however, and hence it is impossible 
to bound the $\delta$-function
interaction in this way without additional help. Our way 
out is to borrow a bit
of kinetic energy (this was also done in Theorem 7.1 
in \cite{LSY94}). So we search for 
\textit{operator inequalities} 
\begin{equation}
a\: p^2 + \frac{1}{Z} \delta(z) \geq w_{Z,a,b}(z)  \label{opiq}
\end{equation}
with appropriate functions $w_{Z,a,b}(z)$, depending on a parameter $b$ in 
addition to $a$ and $Z$ to allow convergence to a delta function.

\begin{lm}[An operator inequality]
\label{lmb2} The inequality (\ref{opiq}) holds for 
\begin{equation}
w_{Z,a,b}(z)=\frac{1}{Z^{2}a}\frac{b^{2}}{(2b+1)}e^{-b|z|/Za}
\end{equation}
\end{lm}

\begin{proof}
With the simple reformulation to 
\begin{equation}
a\:p^2 + \frac{1}{Z} \delta(z) - w_{Z,a,b}(z) \geq 0
\end{equation}
we are on well known territory: The Hamiltonian on the left side
shall have no negative eigenvalue. By the scale transformation
$z \rightarrow Za z$, this inequality is transformed to 
\begin{equation}
p^2 + \delta(z) - W_b(z) \geq 0, \qquad W_b(z) = Z^2\: a\: w_{Z,a,b}(Zaz).
\label{equation:htr}\end{equation}
This inequality will hold for 
\begin{equation}
W_b(z) = \frac{b^2}{2 b + 1} e^{-b|z|}
\end{equation}
if it is true for the larger potential 
\[
\widetilde W_b(z) = \frac{b^2}{(2 b + 1)} \frac{e^{-b|z|}}{(1 -
e^{-b|z|})} , 
\]
because the Hamiltonian in (\ref{equation:htr}) is bounded from below by 
\begin{equation}
p^2 + \delta(z) - \widetilde W_b(z).
\label{htrb}\end{equation}
This Hamiltonian has 
\begin{equation} 
\label{possol}
f(z) = 1 - \frac{1}{2b+1} e^{-b|z|} 
\end{equation} 
as a positive symmetric solution to the Schr\"odinger equation 
- as a differential equation - with zero energy.

Now, if (\ref{htrb}) would have a square integrable ground state 
wave function $g(z)$, this wave function would also be symmetric 
under reflection $z\rightarrow -z$, and the delta-function would 
dictate the same value for $g^{\prime}(z)/g(z)$ as 
it does for $f^{\prime}(z)/f(z)$ at $z = 0_{+}$. 
So the question of the existence of $g(z)$ can be dealt with 
by the methods which are used for proving Sturm's comparison 
theorem: We assume that $g(z)$ exists, with 
negative energy $E$.
The Wronskian $W(z) := f^{\prime}(z)g(z) - g^{\prime}(z)f(z)$ 
is zero at $z = 0_{+}$. Its derivative is determined as 
$W^{\prime}(z) = Ef(z)g(z)$. If $g(z)$ is chosen 
positive, then $W^{\prime}(z)$ is negative, which implies 
that $W(z)$ is negative for $z \geq 0$, and 
$g^{\prime}(z)/g(z) > f^{\prime}(z)/f(z)$. This inequality can be 
integrated to give  $g(z)/g(0) > f(z)/f(0)$, a contradiction 
to the assumption of the square-integrability of g(z).

Therefore we know that the Hamiltonian (\ref{htrb}) has 
no negative eigenvalue. And so the operator inequality holds. 
\end{proof}

The $\{W_b(z)\}$ and hence $\{ Z w_{Z,a,b}(z)\}$ are $\delta$-sequences in
the limit $b \rightarrow \infty$. All these functions are positive 
definite,
and finite at the origin: 
\begin{equation}
w_{Z,a,b}(0) < \frac{b}{2 Z^2\: a}.
\end{equation}

With this tool we can now deduce the lower bound for the many body 
Hamiltonian:

\begin{proposition}
[Lower bound in the mean field limit] If $N,Z\to\infty$ with $\lambda=N/Z$ 
fixed, then
\beq\liminf e(Z,N)/Z^3\geq E^{\rm HS}(\lambda).\eeq
\end{proposition}

\begin{proof}
We use the operator inequality (\ref{opiq}) with $w_Z(z) := w_{Z,a,b}(z)$,
($a$ and $b$ will finally be chosen as appropriate powers of $N$)
to bound $\widehat{h}_{Z,N}$ from below. For each $\delta(z_i - z_j)$ we 
use it twice; one time with $a p_i^2$, and a second time with 
$a p_j^2$. Then we add these inequalities and divide by two:

\begin{eqnarray}
\widehat{h}_{Z,N} &=& \sum_{i=1}^N \left[ \left( 1 - a 
\frac{N-1}{2}\right)p_i^2
- \delta(z_i) \right] + \sum_{i<j} \left[ a\frac{p_i^2+p_j^2}2 + 
\frac{1}{Z} \delta(z_i - z_j)\right]  \nonumber \\
&\geq& \sum_{i=1}^N [\ldots] + \sum_{i<j} w_Z(z_i - z_j).
\label{borrow}\end{eqnarray}

At this point the positive definiteness of $w_Z(z)$ becomes essential. It
implies, that for any real valued integrable function $\sigma (z)$: 
\begin{equation}
\frac{1}{2} \int\!\!\int dz dy \left(N \sigma(z) - \sum_{i=1}^N
\delta(z-z_i)\right) w_Z(z-y) \left(N \sigma(y) - \sum_{i=1}^N
\delta(y-z_j)\right) \geq 0
\end{equation}
Expanding this expression and integrating the delta-functions we get 
\begin{eqnarray}
\sum_{i<j} w_Z(z_i - z_j) \geq \sum_i N \int w_Z(z_i - z) \sigma(
z) dz - \frac{N}{2} w_Z(0)\\-\frac{N^2}{2} \int\!\!\int \sigma(
z) w_Z(z-y) \sigma(y) dz dy .
\nonumber
\end{eqnarray}
Combining this with (\ref{borrow}) gives 
\begin{equation}
\widehat{h}_{Z,N} \geq \sum_{i=1}^N h_i(Z,N,\sigma) - \frac{N^2}{2} 
\int\!\!
\int \sigma(z) w_Z(z-y) \sigma(y) dz dy
\label{lbd}\end{equation}
with the one-particle operators 
\begin{equation}
h_i(Z,N,\sigma) = \left(1 - a \frac{N-1}{2}\right) p_i^2 - \delta(z_i)
+ N(\sigma * w_Z)(z_i) - \frac{1}{2} w_Z(0).
\end{equation}
The parameters are now chosen as 
\[
a = N^{-1-\varepsilon}, \qquad b = N^\varepsilon, \qquad \mbox{with } 0 <
\varepsilon < 1/2. 
\]
The fraction of kinetic energy per particle that we borrowed in 
(\ref{borrow}) then decreases as $%
N^{-\varepsilon}$, and the functions $w_Z(z)$ become 
\begin{equation}
w_Z(z) = \frac{N^{1+\varepsilon}}{Z^2} \frac{N^{2\varepsilon}}
{(2 N^\varepsilon + 1)} e^{-z \cdot N^{1+2\varepsilon}/Z} .
\end{equation}
In the mean field limit $N,Z\to\infty$ with $N/Z=\lambda > 0$ fixed  the sequence $
Z w_Z(z)$ is a $\delta$-sequence, and $w_Z(0) \sim \lambda N^{2\varepsilon}/Z
\rightarrow 0$. If $\sigma (z)$ is smooth with 
$|\sigma^{\prime}(z)|\leq \gamma $, then
$|N(\sigma * w_Z)(z) - \sigma (z)| \leq 2\gamma \lambda ^{2} 
N^{-\varepsilon}$. 
The one particle Hamiltonians $h(Z,N,\sigma)$, with 
smooth $\sigma(z)$, converge as quadratic forms pointwise (i.e., for each test 
function) to 
\begin{equation}
h_{\lambda\sigma} = p^2 - \delta(z) + \lambda \sigma(z) .
\end{equation}
Moreover 
\begin{equation}
\label{nearend}
h(Z,N,\sigma) \geq h_{\lambda\sigma} - (N^{-\varepsilon}/2)p^2 -%
2\delta \lambda ^{2} N^{-\varepsilon} -%
(\lambda ^{2}/2) N^{2\varepsilon -1}.
\end{equation} 
Since the ground state energies of operators of the type 
$\alpha p^2 + V$ are concave functions of $\alpha$ and hence continuous in 
$\alpha$, the ground state energies of the right side 
of (\ref{nearend}) converge in the limit $N \rightarrow \infty$.

The ground state energy of $h_{\lambda \sigma}$ is a concave 
functional $e[\lambda\sigma]$, and the lower bound (\ref{lbd}), 
when divided by the number of electrons $N$, gives 
\begin{equation}\label{I}
\liminf_{N,Z \rightarrow \infty \atop N/Z = \lambda} \frac{1}{N}
\widehat{e}(Z,N) \geq e[\lambda\sigma] - \frac{\lambda }{2} \int 
\sigma^2(z) dz =: \mathcal{I}_\lambda[\sigma].
\end{equation}
Inserting the mean field density $\rho $ for $\lambda \sigma$ 
(i.e., the minimizer of (64) which satisfies Eq.\ (3.8) of \cite{LSY94}) 
gives the mean field energy, divided by $\lambda $, as a 
lower bound to the limit of the energy per electron. 
\end{proof}

We remark that searching for the supremum of $\mathcal{I}_\lambda[\sigma]$ 
in (\ref{I})
also leads to the mean field equation of \cite{LSY94}: Assuming
$e(\lambda\sigma)=\langle\psi,h_{\lambda\sigma}\psi\rangle$ with a normalized 
$\psi$ the variational condition on 
$\sigma(z)$ for maximizing $\mathcal{I}_\lambda[\sigma]$ is 
\begin{equation}
\sigma(z) = \psi^2(z).
\end{equation}
Inserting this into the Schr\"odinger equation 
$h_{\lambda\sigma}\psi=\mu_\lambda\psi$ for $\psi$ gives
\begin{equation}
- \psi^{\prime\prime}(z) - \delta(z) \psi(0) + \lambda \psi^3(z) = -
\mu_\lambda \psi(z),
\end{equation}
i.e., Equation (3.8) in \cite{LSY94}.

Finally we remark that the energy per electron, $\widehat{e}(Z,N)/N$,
approaches the mean field limit monotonously. There is also a 
subadditivity
property, which in the limit becomes concavity of $E^{\mathrm{HS}}(\lambda
)/\lambda $. These properties of the approach to a mean field hold in 
some other cases too, as will be shown elsewhere \cite{B99}.
\section{Conclusions}
We have shown that the energy of an atom in a strong magnetic field $B$ 
approaches, 
after division by $(\ln B)^2$, the energy of a many body Hamiltonian with 
delta 
interactions in one dimension as $B\to\infty$. This delta function model is 
not 
explicitly solvable, but an upper bound to the energy can be given in terms of 
another model with the same mean field limit and 
where we can explicitly calculate the ground state energy. In the latter model 
an atom with nuclear charge $Z$ can bind up to $2Z$ 
electrons. Whether this represents the true state of affairs for the atomic 
Hamiltonian in the $B\to\infty$ limit is an open problem.

\section*{Acknowledgements}
J.P.\ Solovej and J. Yngvason were supported in part by the EU 
TMR-grant FMRX-CT 96-0001. J.P.S. was also supported in part by  MaPhySto -- 
Centre for Mathematical Physics and Stochastics, funded by a grant from The 
Danish National Research Foundation, and by a grant from the Danish Natural 
Science Research Council.

\section*{Appendix} 
We prove here that the ground state energy $\widetilde e(Z,N)$ of the 
Hamiltonian (\ref{newh}) is independent of $N$ if $N\geq 
2Z+1$.

\medskip
\noindent{\bf PROPOSITION (Maximal negative ionization for the comparison 
model).} {\it If $N\geq 2Z+1$, then 
\begin{equation}\widetilde e(Z,N)= \widetilde e(Z,N_{{\rm o}})
\end{equation}
where $N_{{\rm o}}$ is the largest integer strictly smaller than 
$2Z+1$. Moreover, there is then no $L^2$-function with 
$\widetilde e(Z,N)$ as an eigenvalue.}

\medskip
\noindent{\it Proof.} In the cone ${\cal M}$ defined by (\ref{cone}) we 
consider the wave function
\begin{equation}
\check\psi(z_{1},\dots,z_{N})=\prod_{i=1}^{N_{{\rm o}}}
e^{-\kappa _{i}z_{i}}\prod_{j=N_{{\rm o}}+1}^{N}(1-\kappa_{j}z_{j}) , 	
\end{equation}	
with $\kappa_{n}$ defined by (\ref{kappa}). Since $\kappa_{j}\leq 0$ 
for $j\geq N_{\rm o}+1$, the function $\check\psi$ is strictly 
positive.  We extend $\check\psi$ symmetrically from 
${\cal M}$ to all of $\mathbb R^N$ as a continuous function.

The jumps in the logarithmic derivatives of $\check\psi$ at the 
boundary of ${\cal M}$ are not of the right size required
for an eigenfunction of $\widetilde h_{Z,N}$. But $\check\psi$ is an 
eigenfunction of a slightly different operator:
\begin{equation}
\check h_{Z,N}\check\psi=\check e(Z,N)\check\psi	
\end{equation}	
with
\begin{equation}
\check e(Z,N)=-\sum_{i=1}^{N_{\rm o}}\kappa_{i}^2= \widetilde 
e(Z,N_{\rm o})
\end{equation}
and 
\begin{equation}
\check h_{Z,N}=\sum_{i=1}^{N}\left( p_{i}^{2}-\delta (z_{i})\right) 
+%
\frac{1}{2Z}\sum_{i<j}\gamma_{i,j}(z_{1},\dots,z_{N})\left[ \delta (z_{i}-z_{j})+\delta
(z_{i}+z_{j})\right].\label{checkh}
\end{equation}	
with certain functions $\gamma_{i,j}$. 
It is sufficient to specify $\gamma_{i,i+1}(z_{1},\dots,z_{N})$ on the 
boundary of ${\cal M}$ (other cases follow by permutation and/or reflection of the variables) 
and one finds for 
$0\leq z_{1}\leq z_{2}\leq \ldots \leq z_{N}$:
	\beq
	\gamma_{i,i+1}= \left\{ \begin{array}{l@{\quad}l}
	1 & {\rm if}\quad 1\leq i\leq N_{\rm o}-1\\4Z\left(\kappa_{N_{\rm o}}+|\kappa_{N_{\rm 
	o}+1}|(1+|\kappa_{N_{\rm 
	o}+1}|z_{N_{\rm o}})^{-1}\right)& {\rm if}\quad i=N_{\rm 
	o} \\ (1+|\kappa_{i}|z_{i})^{-1}(1+|\kappa_{i+1}|z_{i+1})^{-1}& {\rm if}\quad N_{\rm 
	o}+1\leq i\leq N 
    \end{array}
	\right.\,.
	\eeq	
Since $\gamma_{i,i+1}\leq 1$ for all $i$ one has
\begin{equation}\label{lessthan}
\check h_{Z,N}\leq \widetilde h_{Z,N}. 
\end{equation}
Since $\check\psi$ is stricly positive we can in 
the same way as in (\ref{astara}) write
\begin{equation}
\label{astara2}
	\check h_{Z,N}=\sum _{n}^N\check A_{n}^{*}\check A_{n}+\check e(Z,N)
\end{equation}
with 
\begin{equation}
\check A_{n}=\partial_{z_n}-\partial_{z_{n}}(\ln\check\psi)\,,
\end{equation}
and conclude that $\check e(Z,N)=\widetilde 
e(Z,N_{\rm o})$ is, indeed, the ground state energy 
of $\check h_{Z,N}$. Hence, $\widetilde e(Z,N)=\widetilde e(Z,N_{\rm o})$ for $N\geq 
2Z+1$. 

To see that there are no bound states at the bottom of the spectrum 
of $\widetilde h_{Z,N}$ assume $\psi$ is an eigenfunction to eigenvalue
$\widetilde e(Z,N)$, so that
\begin{equation} 
\langle \psi ,\widetilde{h}_{Z,N} \psi \rangle =
\widetilde{e}(Z,N) \parallel \psi \parallel^{2}. 
\end{equation}
By (\ref{lessthan}) and the equality of the 
ground state energies this implies 
\begin{equation} 
\langle \psi ,\check h_{Z,N} \psi \rangle =
\check e(Z,N) \parallel \psi \parallel^{2}, 
\end{equation}
which, because of (\ref{astara2}),  
is equivalent to the set of differential equations 
\begin{equation} 
\check A_{n} \psi = 0. 
\end{equation} 
These equations have no other solutions than $c \check \psi $, and 
$\check \psi$ is not an $L^2$ function. 
\hfill$\Box$

\end{document}